\documentclass[12pt]{iopart}
\usepackage{iopams,epsf,graphicx}
\begin{document}

\title{The Nernst effect and the boundaries of the Fermi liquid picture}

\author{Kamran Behnia}

\address{Laboratoire Photons et Mati\`ere(UPR5-CNRS),
 ESPCI, 10 Rue Vauquelin, F-75005 Paris, France}

\begin{abstract}
Following the observation of an anomalous Nernst signal in cuprates,
 the Nernst effect was explored in a variety of metals and
 superconductors during the past few years. This paper
reviews the results obtained during this exploration focusing on the
Nernst response of normal quasi-particles as opposed to the one
generated by superconducting vortices or by short-lived Cooper
pairs. Contrary to what has been often assumed, the so-called
Sondheimer cancellation does not imply a negligible Nernst response
in a Fermi liquid. In fact, the amplitude of the Nernst response
measured in various metals in the low-temperature limit is scattered
over six orders of magnitude. According to the data, this amplitude
is roughly set by the ratio of electron mobility to Fermi energy in
agreement with the implications of the semi-classical transport
theory.
\end{abstract}




\section{Introduction}

Nernst effect is the generation of a transverse electric field by a
longitudinal thermal gradient in presence of a finite magnetic
field. It attracted considerable attention during the past few years
following the report on the observation of a finite Nernst effect in
the high-T$_{c}$ cuprate La$_{2-x}$Sr$_{x}$CuO$_{4}$ above the
critical temperature by Ong's group\cite{xu}. Before this report,
Nernst effect had been studied in both conventional\cite{huebener}
and high-T$_{c}$ \cite{palstra,hagen,hagen2,ri} superconductors.
Vortex movement caused by the application of a thermal gradient was
a well-known source of a Nernst signal\cite{huebenerbook}. However,
at least in the community of researchers exploring correlated
electrons, general knowledge regarding other sources of Nernst
effect was fragmentary. In common metals, investigations of the
Nernst effect have been out of fashion since several decades
ago\cite{delves}. As for metals host to correlated electrons, their
Nernst response was still largely unexplored.

This situation has considerably changed during the first years of
this century. The Nernst effect has been studied in a variety of
metals and superconductors. Early measurements of Nernst effect in
the high-T$_{c}$ cuprates performed in the nineties, apart from a
few exceptions\cite{clayhold}, focused on vortex dynamics. These
studies were complemented by new ones probing the anomalous Nernst
response of the pseudogap
state\cite{wang1,wang2,wang3,capan1,capan2,capan3,wen,rullier}. This
has been the subject of a review by Wang, Li and Ong\cite{wang2006}.
Moreover, experiments were performed on other families such as
organic superconductors\cite{wu1,choi,nam1,wu2,nam2} and
heavy-fermion superconductors\cite{bel2,bel3,sheikin,izawa,onose} as
well as on a CDW superconductor such as NbSe$_{2}$\cite{bel1}. In
addition to these studies on clean superconductors, amorphous
superconducting thin films were also explored for the first
time\cite{pourret1,pourret2,spathis}.

One surprising outcome of these explorations was the detection of a
sizeable Nernst signal in many cases in absence of superconductivity
or superconducting fluctuations. This was at first attributed to
exotic physics, since the general scientific opinion was that the
Nernst effect generated by normal quasi-particles in an ordinary
metal should be negligibly small. This line of thought was supported
by the fact that many cases of metals displaying a ``giant Nernst
effect'' displayed other anomalous transport properties. Moreover,
in some cases, such as URu$_{2}$Si$_{2}$\cite{bel3} and
PrFe$_{4}$P$_{12}$\cite{pourret3}, the puzzlingly large Nernst
signal was accompanied by an exotic ground state with an
unidentified order parameter.

The confusion was somewhat dissipated by the rediscovery of
elemental bismuth\cite{behnia1}. The Nernst signal in bismuth is so
large\cite{korenblit,mangez} that it was detectable by late 19th
century's technology\cite{ettingshausen}. A recent study confirmed
the large magnitude of the Nernst effect in bismuth at low
temperatures, which exceeds by more than one order of magnitude the
Nernst response of any correlated metal\cite{behnia1}. Therefore,
the natural question was to check if the semi-classical transport
theory could account for the size of the Nernst effect in bismuth.
This short review argues that the answer to this question is
affirmative. The large Nernst signal in bismuth is a consequence of
a large electron mobility and a small Fermi energy as implied by an
equation first derived by Sondheimer in 1948\cite{sondheimer} and
reformulated recently\cite{oganesyan}. As we will see below, the
available Nernst data for other metals are compatible with this
picture. Thus, the Nernst effect roughly measures the ratio of
electron mobility to Fermi energy in a given metal.

Admitting that normal quasi-particles can generate a sizeable Nernst
signal does not undermine the use of the Nernst effect as a probe of
superconducting fluctuations. On the contrary, knowing the order of
magnitude of the purely metallic Nernst response is indispensable to
disentangle the signal generated by short-lived Cooper pairs (or
eventually short-lived vortices) in the normal state of a
superconductor from the one expected from quasi-particles. Since a
reduced electron mobility and a large Fermi energy lead to a small
Nernst response, an amorphous conventional superconductor such as
Nb$_{x}$Si$_{1-x}$ is an appropriate system for the detection of the
Nernst signal due to short-lived Cooper pairs. Indeed, a finite
Nernst signal was observed in this system in a temperature window
extending up to 30 times T$_{c}$\cite{pourret1,pourret2}. Since this
signal exceeds by three others of magnitude what is expected by the
normal electrons, it is safe to assume that it is not caused by
them. On the other hand, close to T$_{c}$, its magnitude is in
quantitative agreement with what is theoretically expected for
Gaussian fluctuations of the superconducting order
parameter\cite{ussishkin}, leaving little doubt on their origin.

\section{Nernst effect, the semi-classical picture and sign conventions}

The three conductivity tensors, $\overline{\sigma}$, $\overline{\kappa}$ and $\overline{\alpha}$
relate charge current, $\mathbf{J_{e}}$, and heat  current, $\mathbf{J_{q}}$ to electric field,
$\mathbf{E}$ and thermal gradient,$\mathbf{\nabla T}$ vectors:

\begin{equation}
    \mathbf{J_{e}}=\overline{\sigma}. \mathbf{E} - \overline{\alpha}.\mathbf{\nabla T}
\end{equation}
\begin{equation}
    \mathbf{J_{q}}= T \overline{\alpha} . \mathbf{E} - \overline{\kappa}.\mathbf{\nabla T}
\end{equation}

In absence of charge current (i.e. when $\mathbf{J_{e}}=0$ ), the first equation yields:

\begin{equation}
    \mathbf{E} =\overline{\sigma}^{-1}. \overline{\alpha}.\mathbf{\nabla T}
\end{equation}

Therefore, the Nernst signal, N, which is the transverse electric field, $E_{y}$, generated by
a longitudinal thermal gradient,$\nabla_{x}T$, would be:
\begin{equation}
    N= \frac{E_{y}}{\nabla_{x}T} =\frac{\alpha_{xy}\sigma_{xx}-\alpha_{xx}\sigma_{xy}}{\sigma_{xx}^{2}+\sigma_{xy}^{2}}
\end{equation}
\begin{figure}
 \resizebox{!}{0.75\textwidth}{\includegraphics{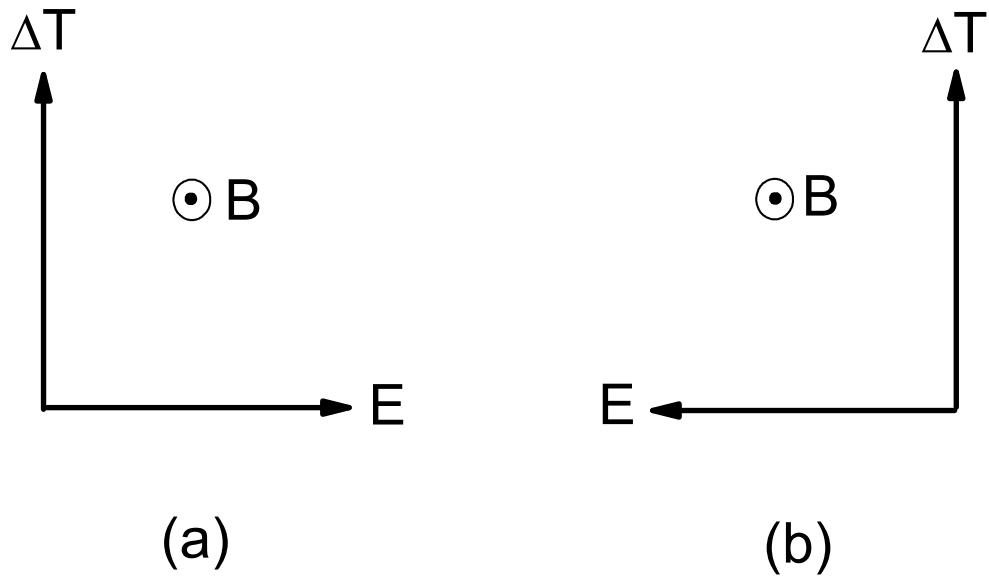}}
\caption{\label{Fig1} The two sign conventions for the Nernst effect : (a) The historical convention and (b) the vortex convention (see text).}
\end{figure}

The sign convention for the Nernst effect has been a source of
confusion, since two different sign conventions have been used (See
Fig. 1). In the first convention, a positive Nernst signal
corresponds to an electric field along the y-axis, when the thermal
gradient is along the x-axis and the magnetic field along the
z-axis. This older convention is the one used in thermoelectric
literature\cite{sugihara} and textbooks\cite{nolas} unrelated to the
Nernst signal of the vortices. According to this convention, the
Nernst signal in bismuth is negative\cite{sugihara,mangez}. However,
a second sign convention has been widely used during the last few
years. According to it, the Nernst signal expected by the vortices
moving from hot to cold is taken as positive\cite{wang2006}. As seen
in Fig. 1, this is opposite of the first convention. The scientific
literature on the superconducting Nernst signal, apart from a few
exceptions\cite{gollnik}, has used this latter convention. The
existence of two opposite sign conventions led to a confusion in the
case of CeCoIn$_{5}$\cite{bel2,onose,izawa}. The Nernst signal in
the normal state of this system is negative according to the first
(or the historical) convention, but positive according to the second
(or the vortex) convention. This feature, not correctly grasped in
the first communication on the observation of the Nernst effect in
this system\cite{bel2} was subsequently corrected\cite{onose,izawa}.
In this text we are going to use the more popular vortex convention,
according to which the Nernst signal in bismuth (which is negative
in the historical convention) would be positive.

The solution of the Boltzmann equation leads to the following link
 between the electric and the thermoelectric conductivity tensors:

\begin{equation}
    \overline{\alpha} =-\frac{\pi^{2}}{3}\frac{k_{B}^{2}T}{e}\frac{\partial\sigma}{\partial\epsilon}|_{\epsilon=\epsilon_F}
\end{equation}

In other words, the thermoelectric response is a measure of the
variation in conductivity caused by an infinitesimal shift in the
chemical potential.

Combining equations 4 and 5 with the definition of the the Hall angle ($\tan\theta_{H}=\frac{\sigma_{xy}}{\sigma_{xx}}$) yields:

\begin{equation}
    N =-\frac{\pi^{2}}{3}\frac{k_{B}^{2}T}{e}\frac{\partial\tan\theta_{H}}{\partial\epsilon}|_{\epsilon=\epsilon_F}
\end{equation}

Such a expression directly linking the Nernst effect to the energy
derivative of the Hall angle was first put forward by Oganesyan and
Ussishkin\cite{oganesyan}. Now, in a one-band picture, the Hall
angle is equal to:
\begin{equation}
   \tan\theta_{H}= \omega_{c}\tau =\frac{e B \tau}{m^{*}}
\end{equation}
Here $\omega_{c}$ is cyclotron frequency, $\tau$ is the scattering
time and $m^{*}$ is the effective mass. Therefore, assuming that the
scattering time is the only energy-dependent component of the Hall
angle, an alternative expression for Eq. 6 would be:
\begin{equation}
    \nu= N/B =-\frac{\pi^{2}}{3}\frac{k_{B}^{2}T}{m^{*}}\frac{\partial\tau}{\partial\epsilon}|_{\epsilon=\epsilon_F}
\end{equation}

This is the expression which appears in Sondheimer's monograph of
1948\cite{sondheimer}. Note the equivalence between equations 6 and
8. A superficial reading of Eq. 8 would erroneously conclude that
the Nernst effect inversely scales with the effective mass. But this
is misleading, since any change in the effective mass would have
consequences on the Fermi energy. For this reason among others, Eq.
6 is more transparent. It states that \emph{if an infinitesimal
shift in the chemical potential, leaves the Hall angle unchanged,
then the Nernst response of the system is nil}. This statement is no
more than one formulation of what has been dubbed Sondheimer
cancelation by Wang and co-workers\cite{wang1}.

There is no well-established experimental procedure to determine
$\frac{\partial\tan\theta_{H}}{\partial\epsilon}|_{\epsilon=\epsilon_F}$.
The simplest approximation is to assume that the Hall angle does not
depend on energy in the vicinity of the Fermi energy. This
assumption would lead to a strictly zero Nernst response.

\section{Two routes towards a finite Nernst signal}

There are two distinct roads to the finite Nernst response observed
in real metals. The first is the presence of multiple bands and the
second is an energy-dependent Hall angle.

\subsection{Ambipolarity}

If
$\frac{\partial\tan\theta_{H}}{\partial\epsilon}|_{\epsilon=\epsilon_F}=
0$, then  Eq. 5 implies the equality:

 \begin{equation}
    \frac{\alpha_{xy}}{\alpha_{xx}}= \frac{\sigma_{xy}}{\sigma_{xx}}
\end{equation}

In a one-band metal, this would lead to a vanishing Nernst response.
However, in a two-band metal with both electron-like and hole-like
carriers, a finite Nernst response can arise even in these
conditions. Indeed, in this case, Eq. 4 should be replaced by:
\begin{equation}
    N=  =\frac{(\alpha^{+}_{xy}+\alpha^{-}_{xy})(\sigma^{+}_{xx}+\sigma^{-}_{xx})-(\alpha^{+}_{xx}+\alpha^{-}_{xx})(\sigma^{+}_{xy}+\sigma^{-}_{xy})}{(\sigma_{xx}^{+}+\sigma_{xx}^{-})^{2}+(\sigma_{xy}^{+}+\sigma_{xy}^{-})^{2}}
\end{equation}
The superscripts + and - refer to hole-like and electron-like bands.
Now, since the signs of $\sigma_{xy}$ and $\alpha_{xx}$ depend on
the sign of the carriers, the validity of Eq. 9 for each band does
not lead to a vanishing numerator in Eq. 10.

\subsection{Energy-dependent mobility}

Even in a one-band picture, the Hall angle can be energy-dependent.
In this case, in a first approximation,
$\frac{\partial\tan\theta_{H}}{\partial\epsilon}|_{\epsilon=\epsilon_F}$
can be replaced by $\frac{\tan\theta_{H}}{\epsilon_F}$. Note that
this is equivalent to assuming that $\tan\theta_{H}$ is a linear
function of energy in the vicinity of the Fermi energy. If the
energy dependence is smooth  but stronger or weaker than linear,
then two expressions should differ by a constant of the order of
unity. The derivative vanishes only in the specific case of the
energy dependence presenting an extremum at the chemical potential.
This would be equivalent to a perfect electron-hole symmetry, a
feature which is often assumed but never demonstrated to occur in
real metals.

Moreover, it is preferable to substitute the Hall angle by what it
physically measures, that is the carrier mobility, $\mu$. The latter
can be expressed as:

 \begin{equation}
    \tan\theta_{H}/B = \mu = \frac{e \tau}{m^{*}}=\frac{e \hbar k_{F}}{\ell_{e}}
\end{equation}
Here $k_{F}$ is the Fermi wavevector and $\ell_{e}$ is the carrier
mean-free-path. This is particularly useful in the case of
multi-band metals. Since the sign of the Hall angle is different for
hole-like and electron-like carriers, the overall Hall angle of an
ambipolar metal can be substantially reduced compared to the Hall
angle [and the mobility] of each band.

These two simplifications lead us to the following expression for
the magnitude of the Nernst coefficient:
\begin{equation}
    \nu=\frac{\pi^{2}}{3}\frac{k_{B}}{e}\frac{K_{B}T}{\epsilon_F} \mu
\end{equation}

Since this expression uses fundamental constants and two
system-dependent parameters, it is easy to use it in order to
confront the measured value of the Nernst coefficient with the
expectations of the semi-classical transport theory. It was first
used in the context of the investigation to the source of the Nernst
signal in URu$_{2}$Si$_{2}$\cite{bel3} and
PrFe$_{4}$P$_{12}$\cite{pourret3}.

Note that the expression proposed by Oganesyan and Ussishkin for a
compensated two-band metal (that is equation A3 in ref.
\cite{oganesyan}):
\begin{equation}
    B \nu=  \frac{2\pi^{2}}{3}\frac{k_{B}}{e}\frac{K_{B}T\tau}{\hbar} \frac{1}{(k_{F}\ell_{B})^{2}}
\end{equation}
is identical to Eq. 10. The strict equivalency between the two
expression becomes explicit if one replaces the magnetic
length,$\ell_{B}$  and the scattering time,$\tau$ by their
corresponding values (that is $\ell_{B}^{2}=\frac{\hbar}{eB}$ and
$\tau=\frac{\ell_{e} m^{*}}{k_{F}}$).

The only difference between the two expressions is the visibility of
the physical parameters which enhance the Nernst response. According
to Eq. 12, the necessary ingredients for an enhanced Nernst signal
is a large electronic mobility and a small Fermi energy. Eq. 13
yields the same message by tracing the source of the Nernst signal
to a long scattering time and a small wave-vector. In the following,
we are going to use Eq. 12, because of the simplicity of
distinguishing three scales: first a universal scale
($\frac{\pi^{2}}{3}\frac{k_{B}}{e} = 283.7 \mu V K^{-1}$ ), second
the inverse of the Fermi energy in Kelvins and finally the mobility
in $T^{-1}$ (or in $m^{2}V^{-1}s^{-1}$). As we shall see below, the
order of magnitude of the available Nernst data is in reasonable
agreement with the expectations of the semi-classical theory.

\subsection{Nernst, Seebeck and Hall coefficients}

When the temperature is much lower than the Fermi temperature, the
Seebeck coefficient of a metallic system,$S$, is expected to become
T-linear with a slop linked to the Fermi temperature through the
following simple expression:

 \begin{equation}
    S= \frac{\pi^{2}}{2} \frac{k_{B}}
    {e}\frac{T}{T_{F}}
\end{equation}

This expression is strictly valid only in the case of a free
electron gas. It is very similar to the one linking the electronic
specific heat, $\gamma$, of a free electron gas to its Fermi
temperature:
\begin{equation}
\gamma = \frac{\pi^2}{2} \frac{k_B}{T_F} n
\end{equation}
Here $n$ is the [molar] carrier density.

Interestingly, the link between $\gamma$ and $S/T$ survives even in
presence of strong electronic interaction. An examination of the
available thermopower and specific heat data in various families of
correlated metals suggests that, at low enough temperatures, the
dimensionless ratio of the Seebeck coefficient to the specific heat
remains of the order of unity \cite{behnia2}.

If in Eq. 12, one replaces the Fermi energy with the Seebeck
coefficient (using Eq. 14) and the mobility with the Hall angle one
finds
\begin{equation}
    \nu  B = \frac{2}{3}S \tan \theta_{H}
\end{equation}

Therefore, Eq. 12 is another formulation of a fundamental link
between the Nernst and Seebeck coefficients through the Hall angle.
However, this equivalency between the two equation only holds in the
case of one-band systems. In a multi-band system, the measured Hall
angle can be significantly lower than the mobility. In bismuth, for
example, in presence of a magnetic field,
$\sigma_{xx}\gg\sigma_{xy}$, but $N \gg S$. Therefore, Eq. 16 fails.
However, as we shall see below, Eq. 12 holds.

\section{Short review of experimental data}
\begin{figure}
 \resizebox{!}{0.75\textwidth}{\includegraphics{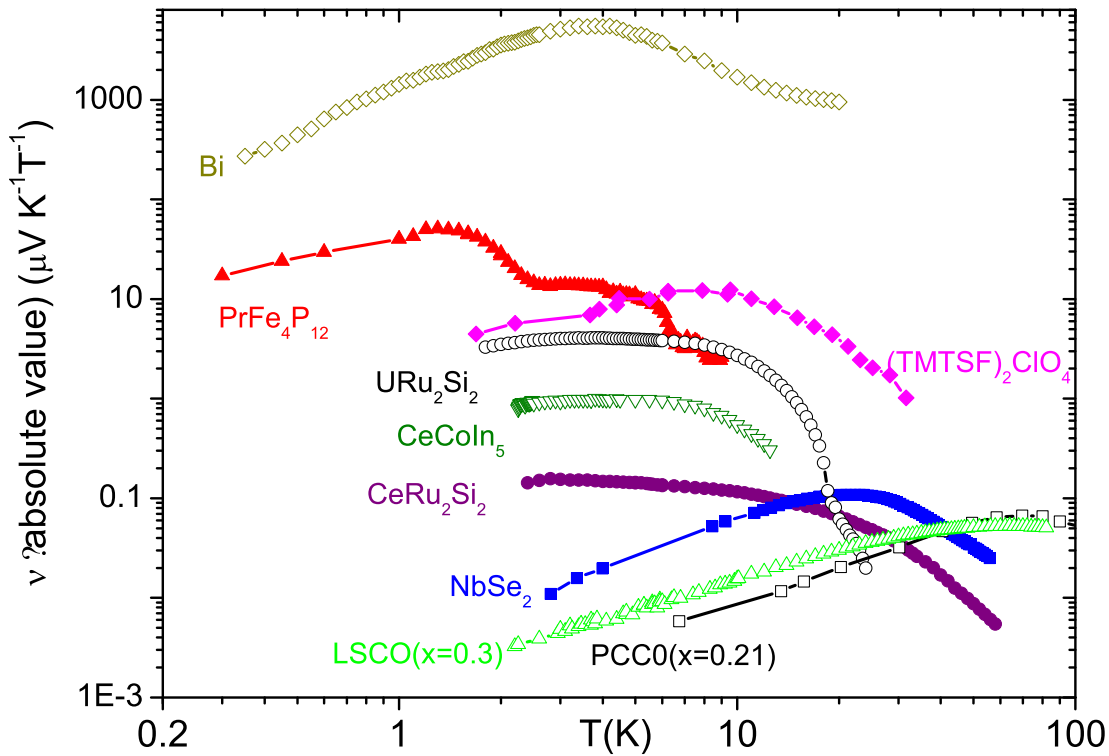}}
\caption{\label{Fig2} The magnitude of the Nernst coefficient in
a number of  metals.}
\end{figure}
In this section we put under scrutiny the available Nernst data. The
study would focus on cases associated with a Nernst effect generated
by normal electrons (as opposed to the signal linked to
superconductivity). The temperature dependence of the absolute value
of the Nernst coefficient of the systems considered in this paper is
presented in Fig.2. Theoretically, the Nernst coefficient should
become T-linear at low enough temperature. It is the order of
magnitude of this T-linear coefficient, which should be confronted
to Eq. 10 (or 11). With this in mind, Fig.3 presents the Nernst data
as a plot of $\nu/T$ vs. temperature. In order to see how the
magnitude of $\nu/T$ at low temperatures conforms to the
expectations of Eq. 10, one needs to extract the magnitude of the
Fermi energy and the electronic mobility in each system. Let us
briefly consider them. A list of extracted parameters is given in
table I.

\subsection{NbSe$_{2}$}

A study of the Nernst effect in NbSe$_{2}$ was reported by Bel and
coworkers\cite{bel1} who attributed the existence of a finite Nernst
signal to ambipolarity. However, the magnitude of the Nernst
coefficient was not put under analysis. NbSe$_{2}$ is not a strongly
correlated system and has a conventional carrier density of about
one carrier per formula unit. Unsurprisingly, its Nernst response is
smaller than correlated metals with high-mobility electrons. We use
the thermoelectric and Hall data to estimate the electronic mobility
and the Fermi temperature. The Fermi temperature can be estimated by
taking the slope of the Seebeck coefficient at low temperature
($S/T=0.3 \mu V K^{-2}$\cite{bel1}) and Eq. 14. The electronic
mobility is estimated using the low-temperature magnitude of the
Hall angle of the sample studied in ref.\cite{bel1}.

\subsection{Ce-based Heavy fermions}

The observation of a large  Nernst coefficient(of the order of $\mu
V K^{-1}T^{-1}$) in CeCoIn$_{5}$ by Bel and co-workers\cite{bel2}
was unexpected. This result was confirmed and extended by subsequent
studies by Onose \emph{et al.}\cite{onose} and Izawa \emph{et
al.}\cite{izawa}. The latter study focused on the low-temperature
region and found that the Nernst response is particularly enhanced
in the vicinity of the field-induced Quantum Critical Point(QCP),
which occurs at 5 T\cite{bianchi,paglione}. We will return to this
study in a following section on quantum criticality. Here, let us
consider the Nernst response in the zero-field limit, which was
probed down to the onset of superconductivity ($T\geq2.2 K$). Since
the transport properties of the system such as thermopower and Hall
coefficient are markedly non-Fermi liquid in the temperature window
extending from T$_c$ to 20K \cite{nakajima}, they cannot be used to
estimate the Fermi temperature and the mobility. One crude estimate
of the Fermi temperature is yielded by the magnitude of the
electronic specific heat, $\gamma$, Eq. 15 and assuming a carrier
density of 1/f.u. (i.e. the system is close to half filling), the
magnitude of $\gamma$ in CeCoIn$_{5}$ (0.6
JK$^{-2}$mol$^{-2}$)\cite{bianchi} implies a Fermi temperature of 60
K.

As for electronic mobility, the value given in table I is extracted
from the Dingle temperature($k_{B}T_{D}= \hbar/ \tau$) and the
effective masses given by de Hass van Alphen
measurements\cite{settai}.

Measurements on CeRu$_{2}$Si$_{2}$ revealed a Nernst signal with a
magnitude somewhat smaller than what was found in
CeCoIn$_{5}$\cite{sheikin}. Here, the zero-field state is a Fermi
liquid and  the Fermi temperature estimated by the slope of
thermopower ($S/T=2.4 \mu VK^{-2}$\cite{amato}) or by the magnitude
of electronic specific
heat($\gamma=0.35Jmol^{-1}K^{-2}$\cite{amato}) are comparable. The
electronic mobility was estimated using the Hall angle
data\cite{bel4}.

\subsection{Heavy-electron metals with unidentified orders}
Among heavy-fermion metals, the magnitude of the Nernst coefficient
in URu$_{2}$Si$_{2}$\cite{bel3} and in Pr-based skutterudite
PrFe$_{4}$P$_{12}$\cite{pourret3} becomes particularly large when
they order. Interestingly, in both these systems the order parameter
appears to be exotic and remains unidentified. However, the large
magnitude of the Nernst coefficient in both cases can be traced back
to the semi-metallic nature of the ordered state implying (in the
language of Eq. 11) a small Fermi wave-vector and a long scattering
time\cite{behnia1}.

In both cases, de Haas-van Alphen studies have detected only small
pockets of Fermi surface. In the case of URu$_{2}$Si$_{2}$, three
frequencies have been detected. The largest correspond to a Fermi
surface whose volume in only 0.02 of the Brillouin
zone\cite{ohkuni}. In the case of PrFe$_{4}$P$_{12}$, only one
frequency is detected, corresponding to 0.0015 of the Brillouin
zone\cite{sugawara}. In both cases, the mass and the volume of the
pockets detected do not sum up to the magnitude of the measured
electronic specific heat. This suggests that one or several massive
low-mobility pockets are not detected yet.

In URu$_{2}$Si$_{2}$\cite{bel3}, the slope of the Seebeck
coefficient continues to increase down to the superconducting
transition temperature ($T_c$ =1.5 K), making it difficult to
extract the zero-temperature value needed to estimate the Fermi
temperature. A more reliable process would be to use
$\gamma=0.065Jmol^{-1}K^{-2}$ and a carrier density of 0.04 per f.u
for this compensated system\cite{kasahara}. This estimate,(T$_F$ =25
K) is comparable to the Fermi temperature of the $\beta$-band
(T$_F$=22 K), which has the lowest Fermi temperature among the three
detected frequencies.

In PrFe$_{4}$P$_{12}$\cite{pourret3}, the situation is more
straightforward: The slope of thermopower ($S/T=56 \mu VK^{-2}$)
yields a rather low Fermi temperature (T$_F$=8K). This value is very
close to the width of the Kondo resonance (8.7 K) found in the
specific heat data\cite{aoki2}.

The mobilities of table I are based on the measurements of the Hall
angle on the crystals which were used in the Nernst studies. Let us
note that in the new generation of ultraclean URu$_{2}$Si$_{2}$
single crystals\cite{kasahara} electronic mobility is at least one
order of magnitude larger. Therefore, according to our current
understanding, the Nernst response of these crystals should also be
enhanced by an order of magnitude.

\begin{figure}
 \resizebox{!}{0.75\textwidth}{\includegraphics{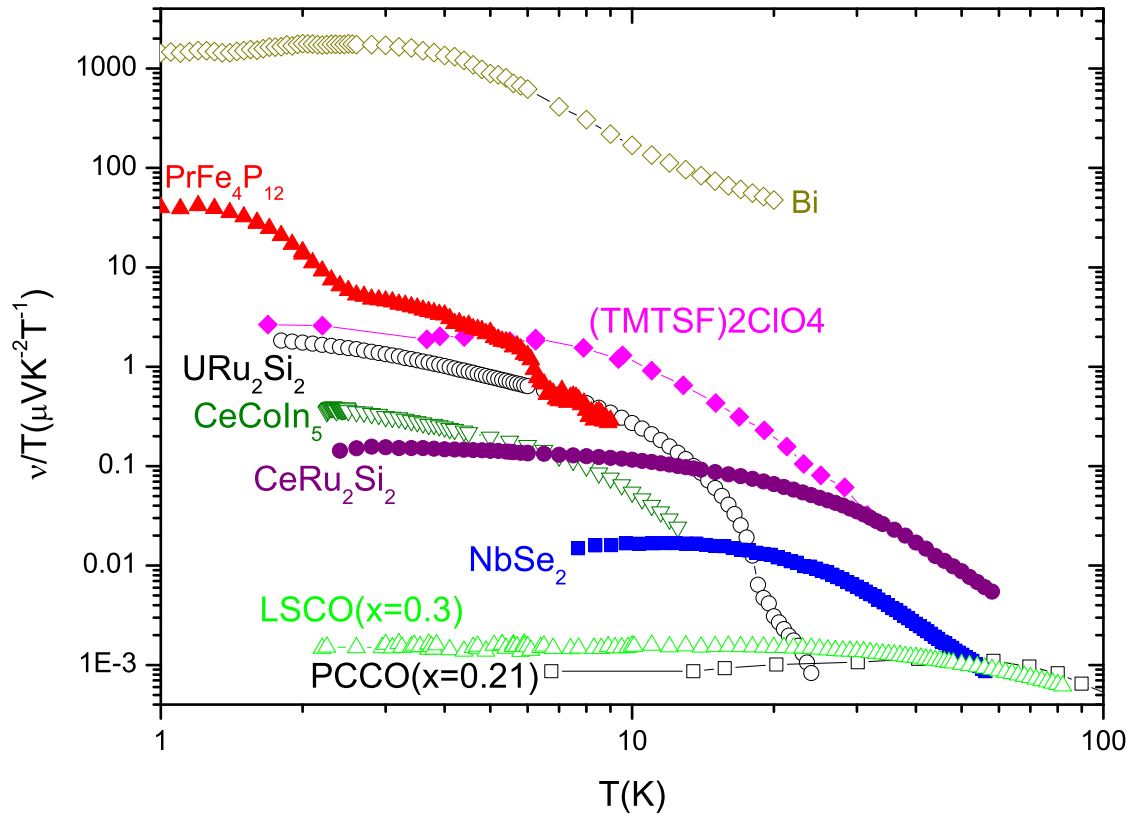}}
\caption{\label{Fig3} The magnitude of the Nernst coefficient divided by temperature.}
\end{figure}

\subsection{Bechgaard salts}
The expression ``giant Nernst effect'' was first employed by Wu
\emph{et al.} following the observation of  a very large resonant
Nernst response in (TMTSF)$_2$PF$_6$\cite{wu1}. A similar feature,
but less pronounced, was observed in the angular dependence of the
Nernst effect in (TMTSF)$_2$ClO$_4$\cite{choi}. In Bechgaard salts
when the field is oriented along some peculiar orientations called
``magic angles'', all transport properties are anomalous and the
large resonant Nernst response is a fascinating problem which should
be addressed in this context\cite{wu2}. The magnitude of the Nernst
coefficient at zero-field on the other hand, may be easier to
understand. Nam \emph{et al.}\cite{nam1} studied the Nernst response
of (TMTSF)$_2$ClO$_4$ in the low field limit and found that the
Nernst coefficient becomes large below the anion-ordering
temperature. Their data are included in Fig. 1 and 2. The slope of
the thermopower in data reported by Nam \emph{et al.}($\sim4 \mu
VK^{-2}$) \cite{nam1} can be used to estimate the Fermi temperature
in (TMTSF)$_2$ClO$_4$. The mobility can be estimated using the
scattering time ($ \tau = 4.3\times 10^{-12}s $) deduced from
angular magnetoresistance studies by  Danner \emph{et
al.}\cite{danner}.

\subsection{Elemental bismuth}
The Nernst effect in elemental bismuth for temperatures exceeding
4.2 K was studied decades ago by three different
groups\cite{korenblit,mangez,sugihara}. The magnitude of the Nernst
response in this semi-metal easily dwarfs all other cases of
``giant'' Nernst effect. The result was rediscovered, confirmed and
extended to lower temperatures recently\cite{behnia1}. The Fermi
surface in bismuth is well-known, the small Fermi temperature for
the electron pocket(27 meV) and hole pocket(11 meV) has been known
for a long time\cite{smith,hartman}. It has also be known that
mobility in bismuth is very large and can exceed
$10^{7}cm^{2}V^{-1}s^{-1}$\cite{hartman}. There is, therefore, no
surprise that bismuth, in which the Nernst effect was originally
discovered, is still the metal with the largest known Nernst
coefficient. The values given in table I correspond to the mobility
of the crystal used in ref.\cite{behnia1} and the Fermi temperature
of the hole pocket (which is more mobile than the electron pocket as
indicated by their respective dingle temperatures\cite{bhargava})
are used. Note that, since bismuth is a compensated system, the Hall
angle is much lower than the mobility of either holes or electrons.
Therefore, Eq. 16 fails.

\begin{table}
 \centering
\begin{tabular}{|c|c|c|c|c|}
  \hline
System & $\nu$/T($\mu$ V K$^{-2} T^{-1}$) & $\mu(T^{-1}$)& $\epsilon_{F} (K^{-1})$& $\frac{\pi^{2}}{3} \frac{k_{B}}{e} \frac{\mu} \epsilon_{F}(\mu V K^{-2} T^{-1})$  \\

\hline
Bi  & 750 & 420 & 130 & 914   \\
\hline
CeRu$_{2}$Si$_{2}$  & 0.16 & 0.2& 180& 0.25  \\
\hline
CeCoIn$_{5}$  & 0.5 & 0.3& 60 & 1.4 \\
\hline
URu$_{2}$Si$_{2}$ & 1.8  &0.08& 25& 0.9  \\
\hline
PrFe$_{4}$P$_{12}$ & 57 & 0.85 & 8 & 30  \\
\hline
(TMTSF)$_{2}$ClO$_{4}$& 2.6 & 0.75& 110 & 1.9 \\
\hline
La$_{1.7}$Sr$_{0.3}$CuO$_{4}$ & 0.0015 &0.01& 5900& $4.8 \times 10^{-4}$  \\
\hline
Pr$_{1.79}$Ce$_{0.21}$CuO$_{4}$ &  9$\times$ 10$^{-4}$ & 0.005  &4300 & $3.3 \times 10^{-4}$ \\
\hline
NbSe$_{2}$ &  0.015 &0.09&1400 & 0.018 \\
\hline
\end{tabular}

\caption{The magnitude of the Nernst coefficient divided by
temperature at low temperature, together with estimations of the
electronic mobility and the Fermi energy in various metals. The
fourth column yields the expected magnitude of $\nu/T$ according to
Eq. 10.}\label{T1}
\end{table}

\subsection{Overdoped cuprates}

It was the discovery of an anomalous Nernst effect in hole-doped
cuprates\cite{xu} which started a tremendous interest in the physics
of the Nernst effect. It is well-known that the existence of a
robust superconducting ground state is a major obstacle to probe the
properties of the normal ground state in the hole-doped cuprates.
This is not the case of the electron-doped ones. Studies of the
Nernst effect in the electron
cuprates\cite{jiang,gollnik,fournier,balci,li1} have detected a
sizeable normal-state Nernst response and attributed it to the
existence of two bands. The data included in Fig.3 and Fig.4 were
those reported for PCCO(x=0.21) by Li and Greene\cite{li1}. At this
doping level, the system is overdoped and its is expected to display
a Fermi liquid behavior. The Fermi temperature and the mobility
given in the table are extracted from the reported data for the
slope of thermopower\cite{li2} and the Hall angle\cite{dagan}.

In the case of hole-doped cuprates, the data presented are
unpublished results obtained in my group on
La$_{1.7}$Sr$_{0.3}$CuO$_4$\cite{jin}. At this doping level,
superconductivity is totally destroyed and no trace of it could be
found down to 0.1 K\cite{nakamae}. The system becomes a Fermi liquid
and its resistivity displays a purely $T^{2}$
behavior\cite{nakamae}. The sign of the Nernst signal is negative in
the vortex convention. The Fermi energy given in table 1 is
calculated using the magnitude of the electronic specific
heat($\gamma=6.9mJmol^{-1}K^{-2}$\cite{nakamae}). The mobility is
extracted from the slope of the B-square
magnetoresistance\cite{nakamae}.

Let us note that the Nernst responses of La$_{1.7}$Sr$_{0.3}$CuO$_4$
and in Pr$_{1.79}$Ce$_{0.21}$CuO$_{4}$ are comparable in magnitude
but opposite in sign. Since the hole-doped cuprate is believed to
have a single large Fermi surface, no ambipolar Nernst effect is
expected there. Its sizeable Nernst response suggests that a
detectable Nernst signal could exist even in the absence of
ambipolarity. In both cases, the low-temperature Nernst signal is
three to four times larger than what is expected according to the
simple picture. This discrepancy appears to be much larger than the
uncertainty on the magnitude of either the mobility or the Fermi
energy and appears significant. It is tempting to link it to a
non-trivial energy-dependence of the relaxation time. Interestingly,
both the magnitude and the temperature-dependence of the Hall number
in La$_{1.7}$Sr$_{0.3}$CuO$_4$differ from what is expected in
isotropic Drude-Boltzmann picture as recently pointed out by
Narduzzo \emph{et al.}\cite{narduzzo}. They have shown that this
departure can be explained by invoking a strong in-plane anisotropic
scattering. It would be interesting to explore the consequences of
such anisotropy for the Nernst response.

\begin{figure}
 \resizebox{!}{0.75\textwidth}{\includegraphics{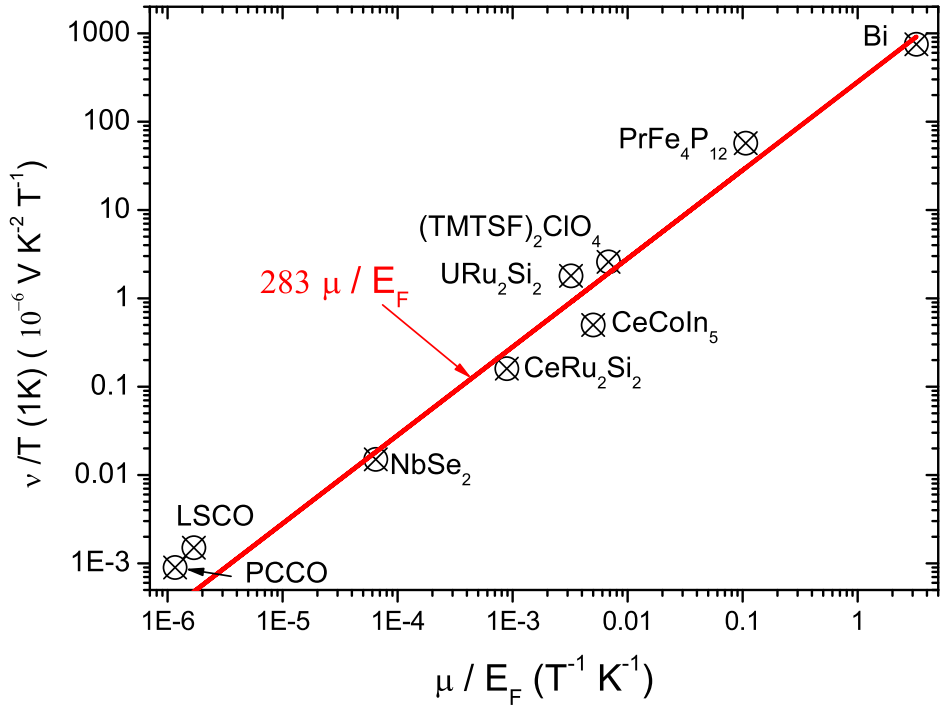}}
\caption{\label{Fig4} The low temperature slope of the Nernst
coefficient as a function of the ratio of electron mobility to the
Fermi energy. The values are those listed in table 1. }
\end{figure}

\section{Overall picture}

Fig.4 displays the magnitude of the low-temperature Nernst
coefficient divided by temperature as a function of the ratio of
mobility divided by Fermi energy. The low-temperature Nernst
coefficient in bismuth is 10$^{6}$ times larger than in
Pr$_{1.79}$Ce$_{0.21}$CuO$_{4}$ and in between lie the scattered
data for other metals. Most of these metals are multi-band systems
with different pockets of Fermi surface and the Fermi energy and the
mobility varies among the bands. Some (such as (TMTSF)$_2$ClO$_4$ )
are very anisotropic with a hierarchy of energy scales, making the
notion of a single Fermi energy questionable. Given all these
uncertainties on the precise magnitude of
$\frac{\mu}{\epsilon_{F}}$, it is remarkable that, as seen in the
figure, the data points scatter around the red line expressing
equation 10 ($\frac{\nu}{T}=283\frac{\mu}{\epsilon_{F}}$).

The main message here is that the order of magnitude of the Nernst
response in the zero-temperature regime is in agreement with the
expectations of the semiclassical theory even when the ratio of
mobility to Fermi energy is changed by six orders of magnitude. Note
also that in several cases the discrepancy points to the
unsatisfactory determination of the Fermi energy. In CeCoIn$_5$, for
example, there is a large uncertainty on the magnitude of the Fermi
energy. In La$_{1.7}$Sr$_{0.3}$CuO$_4$, the Fermi energy extracted
from the slope of thermopower or the coefficient $A$, the prefactor
of inelastic resistivity($\rho=\rho_{0}+AT^{2}$, would yield a
smaller Fermi energy reducing the discrepancy seen in Fig. 4. At
this stage, it is fair to conclude that the ratio of mobility to
Fermi energy is an adequate measure for the expected order of
magnitude of the Nernst response of a Fermi liquid in the
zero-temperature limit.

\section{Nernst effect and quantum criticality}

The Fermi energy, broadly taken, as the main energy scale of the
Fermi liquid vanishes in the vicinity of the Quantum Critical
Point(QCP). If the Nernst response is inversely proportional to the
Fermi energy, it should be enhanced in the vicinity of a QCP. This
has been confirmed by Izawa and co-workers\cite{izawa}, who studied
the thermoelectric response of CeCoIn$_{5}$ in the vicinity of the
field-induced QCP at 5.2 T. The case is well-documented, so that the
relationship between the field-dependence of $\nu/T$ and
$\epsilon_F$ can be checked by looking at other probes of the Fermi
energy, which are the prefactor of T$^2$ resistivity\cite{paglione}
and , $\gamma$, t he electronic specific heat\cite{bianchi}.
Moreover, the direct observation of the Fermi surfaces with de
Haas-van Alphen studies allows to quantitatively link the drop in
the Fermi energy and the mass enhancement of the heaviest detected
band\cite{settai}. The field-dependence of three quantities which
track the Fermi energy, that is $\nu/T$, $A^{1/2}$ and $\gamma$ is
almost identical\cite{izawa}. The available data for the Nernst
coefficient and resistivity suggests that the Fermi energy increases
by a factor of 5 between 16 T and 6 T. [Note that there is no
low-temperature specific heat data for $B > 9T $].

It is interesting to complement these data with the field-dependence
of the zero-temperature Hall coefficient and the effective mass. The
field dependence of the Hall number close to QCP is small(less than
10 percent between 6 T and 8 T)\cite{singh}. Moreover, its magnitude
is comparable with the Hall coefficient in the parent La-based
compound\cite{nakajima}. This suggests that the volume of the large
Fermi surface in CeCoIn$_{5}$ does not change near the QCP. On the
other hand, the variation of the effective mass of the
$\beta_1$-band [as seen by de Haas Van Alphen studies] between 9 T
and 16 T is in quantitative agreement with the decrease in the Fermi
energy implied by the field variation of $\nu/T$ and $A^{1/2}$ (See
Fig. 5). The overall picture drawn by these sets of data is thus the
following: The QCP in CeCoIn$_{5}$ is associated with an exploding
mass without any notable change in the volume of the Fermi surface.
This should present an important constraint for theoretical
scenarios.
\begin{figure}
\resizebox{!}{0.75\textwidth}{\includegraphics{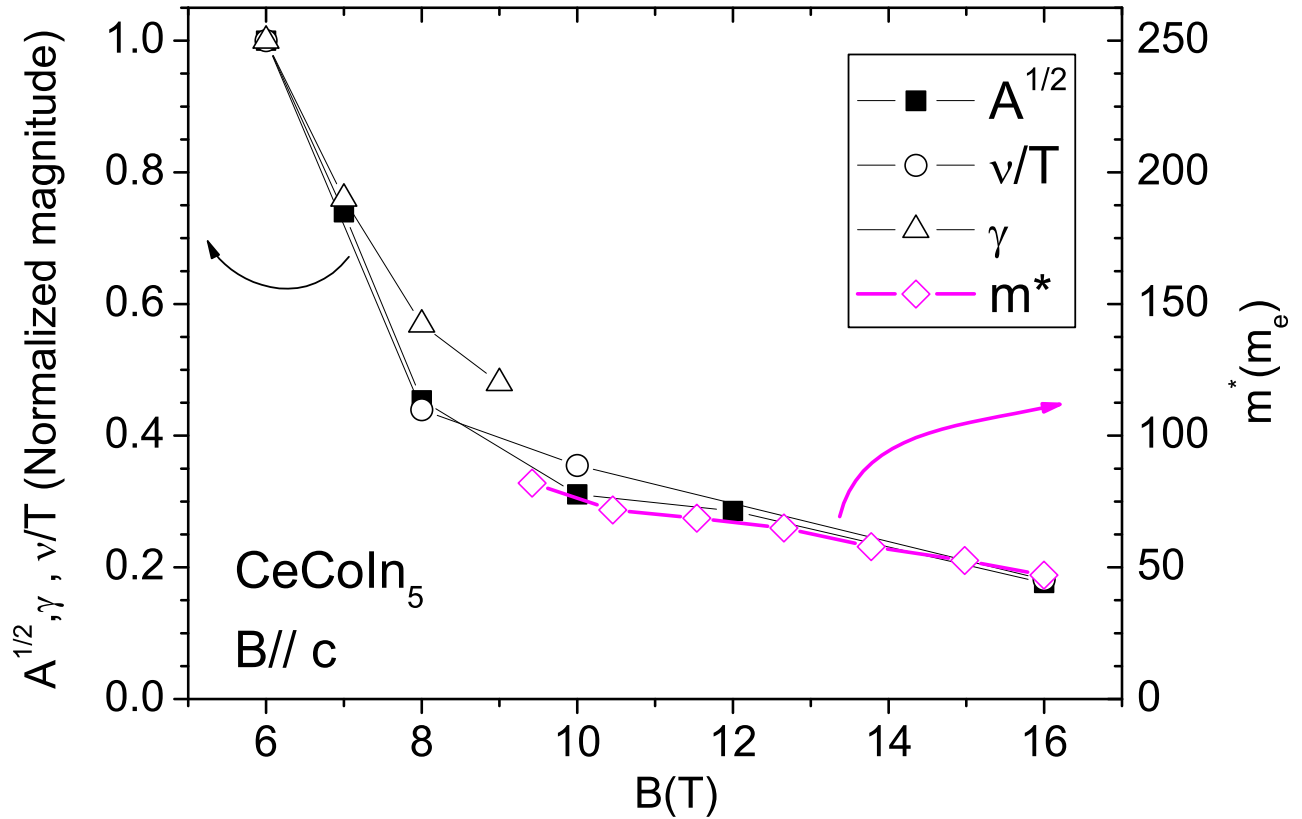}}
\caption{\label{Fig5} Quantum criticality in CeCoIn$_5$. The field
variation of the Nernst coefficient\cite{izawa}, the prefactor of
the T$^2$-resistivity\cite{paglione}, the electronic specific
heat\cite{bianchi} and the effective mass of the $\beta_1$ band
detected by dHvA measurements\cite{settai}. Apart the effective
mass, the magnitude of the other quantities is normalized by their
value at 6T, for the sake of comparison.}
\end{figure}

In the context of a possible link between quantum criticality and an
enhanced Nernst signal, let us note that Li and Greene have reported
a maximum in the doping dependence of the Nernst signal in
Pr$_{1-x}$Ce$_{x}$CuO$_{4}$ close to x=0.15\cite{li2}. Since, this
is close to the critical doping level as evidenced by Hall
measurements\cite{dagan} and backed up by thermopower
data\cite{li1}, it is tempting to speculate on the possible link
between quantum criticality and this maximum. The absence of data
for the critical doping make a definite judgement difficult. Note,
however, that if there is a QCP, it is not driven by the same
mechanism as in CeCoIn$_5$, since it is accompanied with a drastic
change in the Fermi surface topology as suggested by the abrupt
change in the Hall number. Independent of microscopic details, the
link between a maximum in the Nernst response  and a drastic change
in the Hall number is  naturally explained by equation 6. At the
critical doping, the Hall angle is extremely sensitive to any shift
in chemical potential and therefore a maximum in the Nernst response
is expected.

\begin{figure}
\resizebox{!}{0.75\textwidth} {\includegraphics{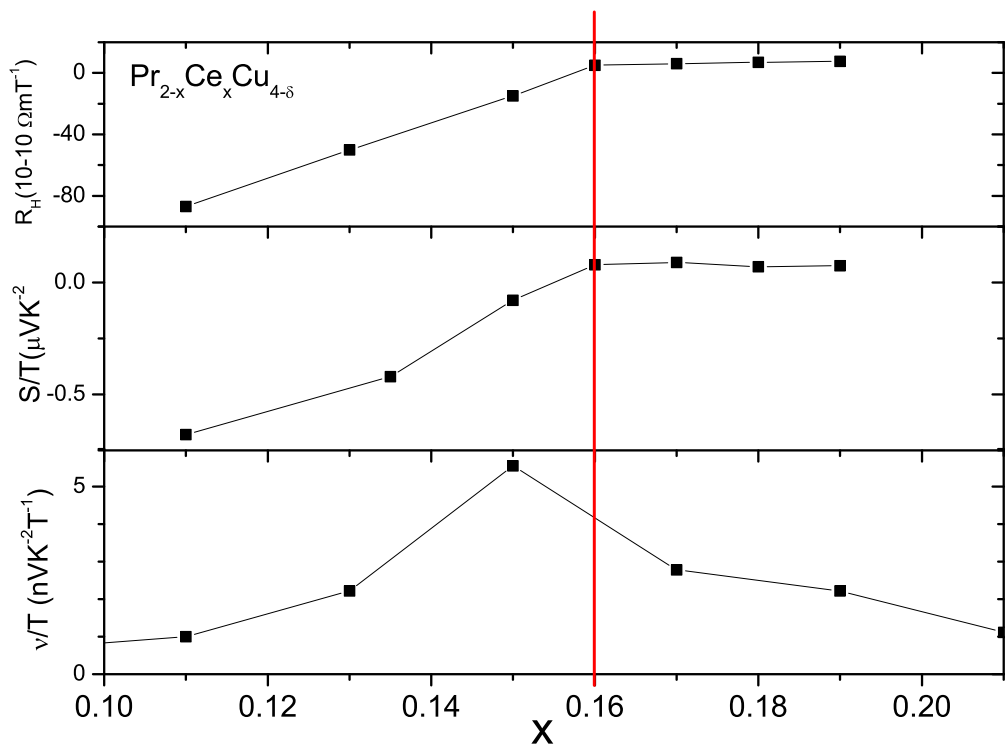}}
\caption{\label{Fig6} Quantum criticality in
Pr$_{1-x}$Ce$_{x}$CuO$_{4}$. The doping dependence of the Hall
coefficient, measured in presence of a field exceeding the upper
critical field and at 0.35 K\cite{dagan}, the low-temperature slope
of the Seebeck coefficient\cite{li1} and the Nernst
coefficient\cite{li2}. The existence of a QCP at x=0.16 was deduced
from the abrupt change in the Hall number and the thermoelectric
response. Nernst coefficient peaks at x=0.15. But the absence of a
data point at x=0.16 makes it difficult to definitely link this peak
to quantum criticality.}
\end{figure}

\section{Conclusions and open questions}

The main conclusion of this short review is that the so-called
Sondheimer cancelation does not imply a zero Nernst signal in a
Fermi liquid. The order of magnitude of the Nernst signal should
become larger as the mobility increases and as the Fermi energy
decreases. This statement is backed by an examination of available
data on different families of remarkable metals. It appears that
this correlation between the magnitude of the Nernst effect and the
Fermi energy remains valid even near a quantum critical point making
the Nernst effect a powerful probe of quantum criticality.

It is important to underline the limits of this simple picture. For
example, the large quantum oscillations of the Nernst effect at low
Landau levels observed in bismuth\cite{behnia3,behnia4} remain
unexplainable in this approach. Explaining why the quantum
oscillations in the Nernst response are much larger than quantum
oscillations of conductivity (the Shubnikov-de Haas effect) remains
a challenge to the theory.

The Nernst effect has proved to be a very powerful probe of
superconducting fluctuations, a subject not addressed by this paper.
While, there is a satisfactory experimental
confirmation\cite{pourret1} of the theory for the Nernst signal of
Gaussian fluctuations\cite{ussishkin}, the issue of the Nernst
effect generated by short-lived vortices (or superconducting phase
fluctuations) in superconductors with small phase stiffness remains
unsettled.

In the underdoped cuprates, even if one assumes that they are Fermi
liquids, the analysis of the Nernst data is particularly complicated
due to the possible role of three distinct sources of the Nernst
signal: normal quasi-particles, short-lived Cooper pairs(amplitude
fluctuations) and short-lived vortices (phase fluctuations).  The
recent detection of a small Fermi surface pocket in underdoped
cuprates\cite{leboeuf} dramatically modifies our picture of the
normal state of these materials. Due to its small Fermi energy, the
size of the Nernst signal generated by this pocket could be
comparable with what is expected due to superconducting
fluctuations.

\section{Acknowledgements}

I would like to thank H. Aubin, R. Bel, C. Capan, J. Flouquet, K.
Izawa, H. Jin, I. Sheikin, Y. Nakajima, Y. Matsuda, M.- A.
M\'easson, A. Pourret , C. Proust and P. Spathis, for their precious
collaboration during the experimental investigation of the Nernst
effect in metals and superconductors in the past few years. I am
grateful to R. Greene for his critical reading of the manuscript. I
would also like to acknowledge the hospitality of the University of
Campinas, Brazil, where this paper was partly written. This work was
supported by Agence Nationale de la Recherche in France and FAPESP
in Brazil.

\end{document}